\documentstyle[12pt]{article}
\begin{document}
\begin {center}
{\bf {\Large $\pi^0\gamma$ invariant mass distribution in the
low energy $ \gamma p \to \omega p $ reaction } }
\end {center}
%\smallskip
%\medskip
\begin {center}
Swapan Das  \\
{\it Nuclear Physics Division,
Bhabha Atomic Research Centre  \\
Mumbai-400085, India }
\end {center}
%\medskip

\begin {abstract}
We study the reaction mechanism for the correlated $\pi^0\gamma$ emission
in the photon induced reaction on the proton target. Since this reaction
is studied in the GeV region, we assume that it proceeds through the
formation of the vector meson in the intermediate state. The vector meson,
being an unstable particle, decays into $\pi^0\gamma$ bosons after
propagating certain distance. Our analysis shows that the  $\pi^0\gamma$
event seen in coincidence in the final state dominantly arises due to the
decay of $\omega$ meson. The calculated results reproduce the measured
$\pi^0\gamma$ invariant mass distribution spectra very well.
\end {abstract}

\section {Introduction}
\label {Int}

~~~~
The vector meson production in the nuclear and particle reactions has
opened up various avenues providing ample opportunities for learning
many interesting topics in physics. The leptoproduction of vector meson
is a potential tool to investigate the hadron structure of photon
\cite{vmsn}. The dilepton production in the intermediate energy region
is undoubtly understood due to the decay and the interference of the
vector mesons \cite{vmsn}. Since the vector meson strongly couples to the
nucleon and its resonances \cite{ptlw,lwf}, the dynamics of these resonances
can be explored by studying the vector meson production process. In fact,
there are predictions of many nucleonic resonances (see Ref.~\cite{oset},
and references their in), which are yet to be found. The vector meson
production could be an useful probe to search these missing resonances.
It should be added that the sub-threshold production of the vector meson
can probe the low lying resonances, like $N(1520)$ \cite{huber}.

In the recent years, the production of vector meson in the nuclear reaction
has drawn considerable attention to explore its properties in the nuclear
medium \cite{vth, vex}. Indeed, it is a fundamental issue in the nuclear
physics. In this context, the $\pi^0\gamma$ invariant mass distribution
spectrum was measured, in the recent past, by the CBELSA/TAPS collaboration
at the electron stretcher accelerator (ELSA) in Bonn \cite{elsa2} to look for
the medium modification on the omega meson in the Nb nucleus. They had also
taken data for the above spectrum due to $\gamma p$ reaction. Here, we
present the reaction mechanism for the $(\gamma,\pi^0\gamma)$ reaction
on the proton target in the GeV region. In this energy region,
the $\pi^0\gamma$ in the final state (according to the particle data group
\cite{pdg1}) can arise due to the decay of the low lying vector mesons,
such as $\rho^0 (768), ~\omega (782) ~\mbox{and} ~\phi (1020)$.  Therefore,
we consider this reaction proceeds through three steps: (i) the formation
of vector mesons, (ii) the propagation of these vector mesons, and (iii)
the decay of these mesons into $\pi^0\gamma$ channel. Symbolically, this
reaction goes as $ \gamma p \to V p ; ~ V \to \pi^0\gamma $, where $V$
stands for the above mentioned vector mesons
$ (i.e., V \equiv \rho^0, ~\omega ~\mbox{and} ~\phi) $.

The qualitative comparison (presented below) amongst the $\rho^0$,
$\omega$ and $\phi$ mesons' contributions to the above reaction shows
that the $\omega$ meson contributes dominantly to this reaction.
The data on the $\gamma p$ reaction show that the production cross section
for the $\rho^0$ meson is the largest compared to those for the other two
vector mesons in the energy region considered here. For example, the measured
cross section at 1.5 GeV beam energy, as reported by various group, are
$ \sigma_t (\gamma p \to \rho^0 p) \simeq    23   \mu $b  \cite{abbc},
$ \sigma_t (\gamma p \to \omega p) \simeq   6.51  \mu $b  \cite{bar1}, and
$ \sigma_t (\gamma p \to \phi p)   \simeq   0.17  \mu $b  \cite{bar2}.
Therefore,
the measured $ \sigma_t (\gamma p \to \rho^0 p) $ is about 3.53 times larger
than the measured $ \sigma_t (\gamma p \to \omega p) $, and the former is
about 135.29 times larger than the measured
$ \sigma_t (\gamma p \to \phi p) $.
The vector meson propagator $G_\omega(m)$ is given by
\begin{equation}
G_V (m) = \frac{ 1 }{ m^2-m^2_V + im_V \Gamma_V(m) },
\label{gvv}
\end{equation}
where $V$ stands for $\rho^0$, $\omega$ and $\phi$ mesons as mentioned
above. For these mesons, we have values for their resonance masses and
total widths \cite{pdg1}:
$ m_{V (\equiv \rho^0)} \simeq 768 $ MeV,
$ \Gamma_{V (\equiv \rho^0)} \simeq 151  $ MeV;
$ m_{V (\equiv \omega)} \simeq 782 $ MeV,
$ \Gamma_{V (\equiv \omega)} \simeq 8.43 $ MeV;
$ m_{V (\equiv \phi)} \simeq 1020 $ MeV,
$ \Gamma_{V (\equiv \phi)} \simeq 4.43 $ MeV.
The data taken at ELSA \cite{elsa2} show the peak at $\sim 780$ MeV in
the $\pi^0\gamma$ invariant mass distribution spectrum. Around
this mass, the propagators for both $\rho^0$ and $\omega$ mesons behave as
$ G_V (m \approx 782 ~\mbox{MeV}) \sim \frac{-i}{m_V\Gamma_V }
~ (V \equiv \rho^0, \omega)$, where as $G_\phi(m)$ goes as
$ G_\phi (m \approx 782 ~\mbox{MeV}) \sim \frac{ 1 }{ m^2-m^2_\phi } $.
The
decay widths for $\rho^0$, $\omega$ and $\phi$ mesons in the
$\pi^0\gamma$ branch (according to the Ref.~\cite{pdg1}) are
$ \Gamma(768)_{\rho^0 \to \pi^0\gamma} \approx 0.12 $ MeV,
$ \Gamma(782)_{\omega \to \pi^0\gamma} \approx 0.72 $ MeV, and
$ \Gamma(1020)_{\phi \to \pi^0\gamma}  \approx 0.006 $ MeV.
In fact, $ \Gamma_{\rho^0 \to \pi^0\gamma} (m) $ is negligibly larger than
0.12 MeV and $ \Gamma_{\phi \to \pi^0\gamma} (m) $ could be much less than
0.006 MeV at $m \approx 782$ MeV.
Therefore,
at 1.5 GeV the ratios $ \frac{\sigma_t (\gamma p \to \omega p)}
{\sigma_t (\gamma p \to \phi p)} \simeq 38.29 $,
$ | \frac{ G_\omega (m \approx 782) }{ G_\phi (m \approx 782) } |^2
\sim 4.2 \times 10^3 $,
and
$ \frac{\Gamma(m \approx 782)_{\omega \to \pi^0\gamma}}
       {\Gamma(m \approx 782)_{\phi \to \pi^0\gamma}} > 120 $ show that we
can safely ignore the contribution to the cross section originating from the
$\phi$ meson of mass around 782 MeV.
For $\rho^0$ and $\omega$ mesons, the above data show 
$ \frac {\sigma_t (\gamma p \to \omega p)}
        {\sigma_t (\gamma p \to \rho^0 p)} \approx 0.28 $,
$ | \frac { G_\omega (m \approx 782) }{ G_{\rho^0} (m \approx 782) } |^2
    \sim 3.1 \times 10^2 $, and
$ \frac{\Gamma(m=782)_{\omega \to \pi^0\gamma}}
{\Gamma(m=782)_{\rho^0 \to \pi^0\gamma}}  \approx  6 $.
Therefore, this analysis illustrates qualitatively that the contribution to
the $\pi^0\gamma$ emission from the $\omega$ meson decay is a factor about
521 times larger than that originating due to the $\rho^0$ meson decay.
Similar analysis shows this factor is about $10^3$ at
$E_\gamma$ equal to 1.2 GeV.

There were two aspects in the experiment (done at ELSA \cite{elsa2}) to be
mentioned. One aspect in this measurement was the gamma beam of wide
energy spread ($0.64-2.53$ GeV), since it was the tagged photon produced
by the bremsstrahlung radiation of the 2.8 GeV electron on the Pb target.
This is unlike to the conventional experiments where the beam energy is
used to be taken almost monoenergetic. In fact, the measured $\pi^0\gamma$
invariant mass distribution spectrum has been reported for definite $\omega$
meson momentum bin instead of a particular incident energy. The another
aspect of this experimental
set-up was the large width in the detecting system (55 MeV), which is about
6.5 times larger than the width of the $\omega$ meson (i.e., $ \Gamma_\omega
(m=782 ~\mbox{MeV}) = 8.43 $ MeV in the free state) produced in the
$\gamma p$ reaction. Therefore, the measured $\pi^0\gamma$ invariant mass
distribution spectrum showing its width about 55 MeV can be believed
undoubtly due to the width of the detector resolution. To compare the
calculated results with the data, we incorporate these two aspects in our
formalism presented in sec.~2. The results of this study are discussed in
sec.~3, and we give the conclusion in sec.~4 in this manuscript.

\section {Formalism}
\label {for}

~~~~
The formalism for the $\pi^0\gamma$ emission in the $\gamma p$ reaction,
as mentioned above, consists of the production, propagation, and the decay
of the omega meson produced in the intermediate state, i.e.,
$ \gamma p \to \omega p^\prime; ~ \omega \to \pi^0\gamma^\prime $. The primes
on $p$ and $\gamma$ are used at present to distinguish them from the initial
state particles, which have been dropped afterward for convenience. The
T-matrix $T_{fi}$ for this process can be written as
\begin{equation}
T_{fi} = ( \pi^0, \gamma^\prime | \Gamma_{\omega\pi\gamma} | \omega )
            G_\omega (m) F ( \gamma p \to \omega p^\prime ),
\label{tmx}
\end{equation}
where $G_\omega (m)$ denotes the propagator for the $\omega$ meson.
$ ( \pi^0\gamma^\prime | \Gamma_{\omega\pi\gamma} | \omega ) $ in this
equation is the matrix element for $ \omega \to \pi^0\gamma^\prime $
due to intrinsic coordinates. It is governed by the Lagrangian
density \cite{dazi}:
\begin{equation}
{\cal L}_{\omega\pi^0\gamma} = \frac{ f_{\omega\pi\gamma} }{ m_\pi }
\epsilon_{\mu\nu\rho\sigma} \partial^\mu A^\nu \pi^0 \partial^\rho
\omega^\sigma,
\label{lag}
\end{equation}
where $A$ appearing in this equation represents the photon field.
$f_{\omega\pi\gamma}$ is the constant for the $\omega\pi\gamma$ coupling.
It is equal to 0.095, as extracted from the width of
$\omega \to \pi^0\gamma$ \cite{pdg1}.

In Eq.~(\ref{tmx}), $ F (\gamma p \to \omega p^\prime) $ describes the
production mechanism for the $\omega$ meson in the $\gamma p$ reaction.
It is given by
\begin{equation}
F (\gamma p \to \omega p^\prime)
= -4\pi E_\omega
\left [ \frac{1}{E_\omega} + \frac{1}{E_{p^\prime}} \right ]
< \omega, p^\prime | f_{\gamma p \to \omega p^\prime} (0) | \gamma, p >,
\label{fgo1}
\end{equation}
where $ f_{\gamma p \to \omega p^\prime} (0)$ is the forward amplitude for
the $ \gamma p \to \omega p^\prime $ reaction.

The differential cross section for the reaction described above can be
written as
\begin{equation}
d\sigma
= \frac{1}{2E_\gamma} (2\pi)^4 \delta^4 (k_i-k_f) <|T_{fi}|^2>
  \frac{m_p}{E_{p^\prime}}\frac{d^3{\bf k}_{p^\prime}}{(2\pi)^3}
  \frac{1}{2E_{\pi^0}}\frac{d^3{\bf k}_{\pi^0}}{(2\pi)^3}
  \frac{1}{2E_{\gamma^\prime}}\frac{d^3{\bf k}_{\gamma^\prime}}{(2\pi)^3}.
\label{ds1}
\end{equation}
The annular bracket around the $T_{fi}$-matrix indicates
the average over the polarization and spin in the initial state and the
summation over the polarization and spin in the final state. Since the
$\pi^0$ and $\gamma$ bosons in the final state are considered originating
due to the decay of the $\omega$ meson, the above expression can be worked
out in terms of the $\omega$ meson mass $m$ (i.e., the $\pi^0\gamma^\prime$
invariant mass) as
\begin{equation}
\frac{d\sigma (m,E_\gamma)}{dm}
= \int d\Omega_\omega [KF] \Gamma_{\omega \to \pi^0\gamma^\prime}(m)
  |G_\omega (m)|^2 | F(\gamma p \to \omega p^\prime) |^2.
\label{dsc1}
\end{equation}
Where $d\Omega_\omega$ is the infinitesimal solid angle subtended by the
$\omega$ meson momentum $ {\bf k_\omega ( =k_{\pi^0}+k_{\gamma^\prime}) } $.
$ \Gamma_{\omega \to \pi^0\gamma^\prime}(m) $ denotes the width for the
$\omega$ meson of mass $m$ decaying at rest into $\pi^0\gamma$ channel. It
is expressed in Eq.~(\ref{wdth2}). $[KF]$ represents the kinematical factor
for this reaction, which is given by
\begin{equation}
[KF] = \frac{3\pi^3}{(2\pi)^6} \frac{ k^2_\omega m_p m^2 }
{ k_\gamma | k_\omega (E_\gamma+m_p) - {\bf k_\gamma . {\hat k}_\omega}
E_\omega |}.
\label{kf}
\end{equation}

The Eq.~(\ref{dsc1}) illustrates the differential cross section for the
$\omega$ meson mass distribution due to fixed beam (gamma) energy
$E_\gamma$. But the measurement was done by the CBELSA/TAPS
collaboration \cite{elsa2}, as mentioned earlier, using a range of incident
$\gamma$ energy instead of fixed beam energy. We incorporate it in our
calculation by modulating the cross section in Eq.~(\ref{dsc1}) with the
beam profile function $W(E_\gamma)$ \cite{kho}, i.e.,
\begin{equation}
\frac{d\sigma (m)}{dm} = \int^{E_\gamma^{mx}}_{E_\gamma^{mn}}
dE_\gamma W(E_\gamma) \frac{d\sigma (m,E_\gamma)} {dm}. 
\label{dsc2}
\end{equation}
$E_\gamma^{mn}$ and $E_\gamma^{mx}$ are equal to 0.64 GeV and 2.53 GeV
respectively, as mentioned in Ref.~\cite{elsa2}. The profile function
$W(E_\gamma)$ for the $\gamma$ beam, originating due to bremsstrahlung
radiation of the electron, varies as
$W(E_\gamma) \propto \frac{1}{E_\gamma}$ \cite{kho}.

The calculated cross section due to Eq.~(\ref{dsc2}) can't reproduce the
measured distribution, since the detecting system in the experimental set-up
(as mentioned earlier) had large resolution width, i.e., 55 MeV, \cite{elsa2}.
This issue is incorporated in the formalism by folding the differential
cross section in Eq.~(\ref{dsc2}) with a Gaussian function $R(m,m^\prime)$:
\begin{equation}
\frac{d\sigma (m)}{dm}
= \int dm^\prime R(m,m^\prime) \frac{d\sigma (m^\prime)}{dm^\prime}.
\label{dsc3}
\end{equation}
The function $R(m,m^\prime)$ accounts the resolution (or response) for the
detector. The expression for it \cite{knoll} is
\begin{equation}
R(m,m^\prime)
= \frac{1}{\sigma\sqrt{2\pi}} e^{-\frac{(m-m^\prime)^2}{2\sigma^2}}.
\label{resl}
\end{equation}
Here, $\sigma$ is related to the full width at the half-maxima (FWHM) of this
function as FWHM$=2.35\sigma$. We take the value for FWHM equal to 55 MeV,
so that the function $R(m,m^\prime)$ can describe properly the resolution
for the detector used at ELSA \cite{elsa2}.

\section {Results and Discussion}
\label {rlds}
~~~~
The amplitude for the $ \gamma p \to \omega p $ reaction, i.e.,
$ f_{\gamma p \to \omega p} (0) $ needed in Eq.~(\ref{fgo1}), is
related to the four momentum $q^2$ transfer distribution
$ d\sigma (\gamma p \to \omega p) / dq^2 $ \cite{shmt} as
\begin{equation}
| f_{\gamma p \to \omega p} (q^2) |^2
= \frac{k^2_\gamma}{\pi}
  \frac{d\sigma}{dq^2} (\gamma p \to \omega p).
\label{fgo2}
\end{equation}
Therefore, the energy dependent values for 
$ | f_{\gamma p \to \omega p} (0)|^2 $ (required to calculate the cross
sections in Eqs.~(\ref{dsc2}) and (\ref{dsc3})) can be extracted from the
four momentum transfer distribution
$ d\sigma (\gamma p \to \omega p) / dq^2 $. In fact, the forward
$ d\sigma (\gamma p \to \omega p) / dq^2 $ is obtained by
extrapolating the measured $ d\sigma (\gamma p \to \omega p) / dq^2 $
to $q^2=0$ for $ E_\gamma \ge 1.6 $ GeV \cite{shmt, stk}.
It
should be mentioned that the present study dominates over the $\omega$
meson production in the $\gamma p$ reaction at lower energy, i.e.,
$ E_\gamma \le 1.6 $ GeV. Recently, the experiment had been carried out
on the low energy $ \gamma p \to \omega p $ reaction with SAPHIR detector
at electron stretcher ring (ELSA), Bonn \cite{bar1}. In this measurement,
they have reported the measured
$ d\sigma (\gamma p \to \omega p) / dq^2 $ vs $ | q^2 - q^2_{min} | $
($q^2_{min}$ is defined in Ref.~\cite{shmt, tmn}) in the energy region
$ E_\gamma = 1.1 - 2.6 $ GeV. In addition, they have also shown that the
measured $ d\sigma (\gamma p \to \omega p) / dq^2 $ behaves as
$ \sim exp [ b(q^2-q^2_{min}) ] $ in the low four momentum $q^2$ transfer
region. Since this energy region is well accord with our study, we
extract $ | f_{\gamma p \to \omega p} (0)|^2 $ from the SAPHIR data and use
it in our calculation.

The $\omega$ meson propagator $G_\omega(m)$ in Eq.~(\ref{dsc1})
can be described by the Eq.~(\ref{gvv}). The total width
$ \Gamma_\omega (m) $ appearing in it is composed of widths due
to omega meson decaying into various channels \cite{pdg1}:
\begin{equation}
\Gamma_\omega \approx
  \Gamma_{\omega \to \pi^+\pi^-\pi^0} (88.8\%)
+ \Gamma_{\omega \to \pi^0\gamma} (8.5\%)
+ \Gamma_{\omega \to \pi^+\pi^-} (2.21\%)
+ \Gamma_{\omega \to l^+l^-} (\sim 10^{-4}\%).
\label{wdths}
\end{equation}
We ignore here widths due to the leptonic decay
$(\Gamma_{\omega \to l^+l^-})$, since they are insignificant. The typical
magnitudes for them are of the order of keV or less, where as the widths
for the $\omega$ meson decaying into other channels in Eq.~(\ref{wdths})
are within the range of 100 keV to 10 MeV.

$ \Gamma_{\omega \to \pi^+\pi^-\pi^0} (m) $ in the above equation represents
the width for the $ \omega \to \pi^+\pi^-\pi^0 $ channel. We use the form
for it as derived by Sakurai \cite{saku}:
\begin{equation}
\Gamma_{\omega \to \pi^+\pi^-\pi^0} (m)
=\Gamma_{\omega \to \pi^+\pi^-\pi^0} (m_\omega)
 \frac{m}{m_\omega} \frac{(m-3m_\pi)^4}{(m_\omega-3m_\pi)^4}
 \frac{U(m)}{U(m_\omega)},
\label{wdth1}
\end{equation}
with $ \Gamma_{\omega \to \pi^+\pi^-\pi^0} (m_\omega = 782 ~\mbox{MeV})
\approx 7.49 $ MeV. $ U(m) \to 1 $ as $ m \to 3m_\pi $ and
$ U(m) \to 1.6 $ as $ m \to 787 $ MeV. We have also taken $U(m)$ equal to
1.6 for $ m > 787 $ MeV.

The width $ \Gamma_{\omega \to \pi^0\gamma} (m) $ in Eq.~(\ref{wdths})
arises due to $ \omega \to \pi^0\gamma $ channel. Using the Lagrangian
density $ {\cal L_{\omega\pi\gamma}} $ given in
Eq.~(\ref{lag}), it is evaluated as
\begin{equation}
\Gamma_{\omega \to \pi^0\gamma} (m)
=\Gamma_{\omega \to \pi^0\gamma} (m_\omega)
 \left [ \frac{k(m)}{k(m_\omega)} \right ]^3,
\label{wdth2}
\end{equation}
with $ \Gamma_{\omega \to \pi^0\gamma} (m_\omega) \approx 0.72$ MeV at
$m_\omega = 782$ MeV. $k(m)$ denotes the momentum of pion originating due
to the $\omega$ meson of mass $m$ decaying at rest.

In Eq.~(\ref{wdths})$, \Gamma_{\omega \to \pi^+\pi^-} (m) $ denotes the
width for the $\omega$ meson decaying into $\pi^+\pi^-$ channel. This
channel arises due to the small pure isovector $\rho$ meson component
present in the physical $\omega$ meson \cite{vmsn}. Using the Lagrangian
density ${\cal L}_{\omega\pi\pi}
= f_{\omega \pi\pi} ( \vec {\pi} \times \partial_\mu \vec {\pi}) \cdot
{\bf \omega}^\mu $, the width for this channel is
worked out as
\begin{equation}
\Gamma_{\omega \to \pi^+\pi^-} (m)
=\Gamma_{\omega \to \pi^+\pi^-} (m_\omega) \frac{m_\omega}{m}
 \left [ \frac{k(m)}{k(m_\omega)} \right ]^3.
\label{wdth3}
\end{equation}
The value for $ \Gamma_{\omega \to \pi^+\pi^-}
(m_\omega = 782 ~\mbox{MeV}) $, according to Ref.~\cite{pdg1}, is
approximately equal to 0.19 MeV. $k(m)$ represents the pion momentum in the
$\pi^+\pi^-$ cm of system.

We calculate the cross sections for the $\omega$ meson mass distribution
in the $\gamma$ induced reaction on the proton target. The calculated results
have been compared with the measured $\pi^0\gamma$ invariant mass
distribution spectra in the $ p(\gamma, \pi^0\gamma)p $ reaction, since
the $\pi^0$ and $\gamma$ in the final state (as mentioned earlier) are
assumed to originate due to the decay of the $\omega$ meson produced in the
intermediate state. The measured $\pi^0\gamma$ invariant mass distribution
spectra have been reported in the Ref.~\cite{elsa2} for four $\omega$ meson
momentum bins:
(i)   $0.2 < k_\omega \mbox{(GeV/c)} < 0.4$,
(ii)  $0.4 < k_\omega \mbox{(GeV/c)} < 0.6$,
(iii) $0.6 < k_\omega \mbox{(GeV/c)} < 1$, and
(iv)  $1   < k_\omega \mbox{(GeV/c)} < 1.4$.
Therefore, we have calculated cross section for 
(i) $ k_\omega = 0.21 - 0.39 ~\mbox{GeV/c} $;
(ii)  $ k_\omega = 0.41 - 0.59 ~\mbox{GeV/c} $;
(iii) $ k_\omega = 0.61 - 0.99 ~\mbox{GeV/c} $;
and (iv)  $ k_\omega = 1.01 - 1.39 ~\mbox{GeV/c} $.

The $\omega$ meson mass distribution spectra calculated using the
Eq.~(\ref{dsc2}) are presented in Fig.~1 by the dash-dot curves. The sharp
peak at the mass around 780 MeV and width about 8.43 MeV, appearing in
these curves, are the characteristics features for the $\omega$ meson
produced in the free state. The solid curves in this figure illustrate
the $\omega$ meson mass distribution spectra calculated using the
Eq.~(\ref{dsc3}). These curves arise due to the folding of the calculated
cross section given in Eq.~(\ref{dsc2}) with the detector resolution function
$R(m,m^\prime)$ in Eq.~(\ref{resl}).
Therefore,
the incorporation of the detector resolution function in the calculated
cross section, as shown in this figure, has widen all spectra and
simultaneously, it reduces the cross section at the peak. The enhancement
in the width from $\sim 8.43$ MeV (dash-dot curves) to $\sim 55$ MeV (solid
curves) occurs due to the width appearing in the detector resolution
function $R(m,m^\prime)$, see Eq.~(\ref{resl}).

In Fig.~2, we compare the calculated results due to Eq.~(\ref{dsc3}) with
the measured $\pi^0\gamma$ invariant mass distribution spectra for all
$\omega$ meson momentum bins. This figure shows that the calculated $\omega$
meson mass distribution folded duly with the detector resolution function,
as described above, reproduce the data very well for all bins of the
$\omega$ meson momentum.

\section {Conclusion}
\label {cncl}
~~~~
We have calculated the differential cross section for the $\pi^0\gamma$
invariant mass
distribution in the $\gamma p$ reaction. Our analysis shows that the
$\pi^0\gamma$ appearing in the final state is occurring dominantly
due to the decay of the $\omega$ meson produced in the intermediate state.
The calculated results showing sharp and narrow peak characterize the 
production of the $\omega$ meson in the free state. The incorporation of
the Gaussian function (to describe the detector resolution) in the
calculation broadens the $\omega$ meson mass distribution spectrum, which
is well accord with the data.

\section {Acknowledgement}
\label {ackn}
~~~~
I gratefully acknowledge L.M. Pant for making me aware about the measurement
on the omega meson mass distribution at ELSA. The discussion with D.R.
Chakrabarty on the detector resolution is highly appreciated. The
communication made with E. Oset regarding the beam profile function is
very helpful. I acknowledge D. Trnka and V. Metag for sending the data. 
I thank A.K. Mohanty, R.K. Choudhury and S. Kailas for their support.

{\bf Figure Captions}
\begin{enumerate}

\item
The $\omega$ meson mass distribution spectra are presented for various
$\omega$ meson momentum bins. The dash-dot curves correspond to the
calculated results due to Eq.~(\ref{dsc2}), where as the solid curves
arise due to the Eq.~(\ref{dsc3}) (see text).

\item
The calculated $\omega$ meson mass distribution spectra for various
$\omega$ meson momentum bins are compared with the data (see text). The
solid curves correspond to the calculated results due to Eq.~(\ref{dsc3}).
The histograms represent the measured counts for the $\pi^0\gamma$
invariant mass distribution spectra \cite{elsa2}, normalised to the
respective calculated peaks.

\end{enumerate}

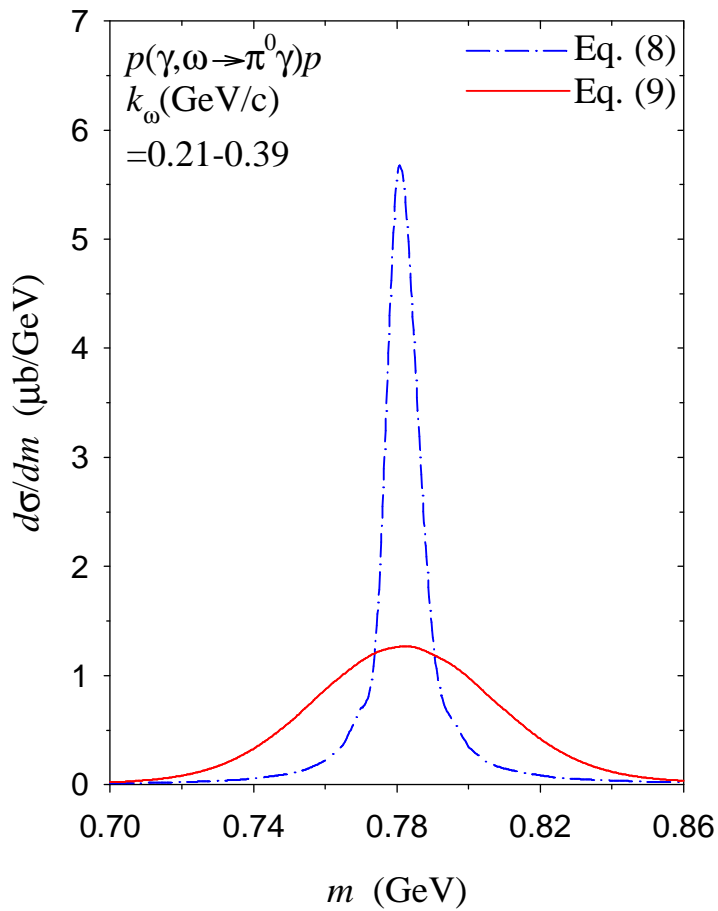
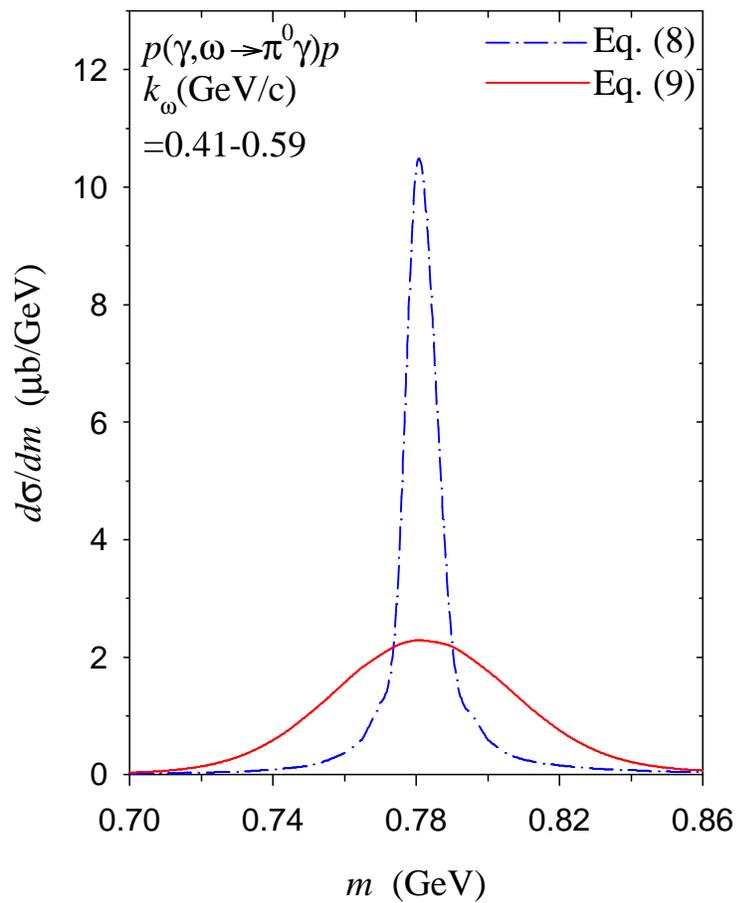
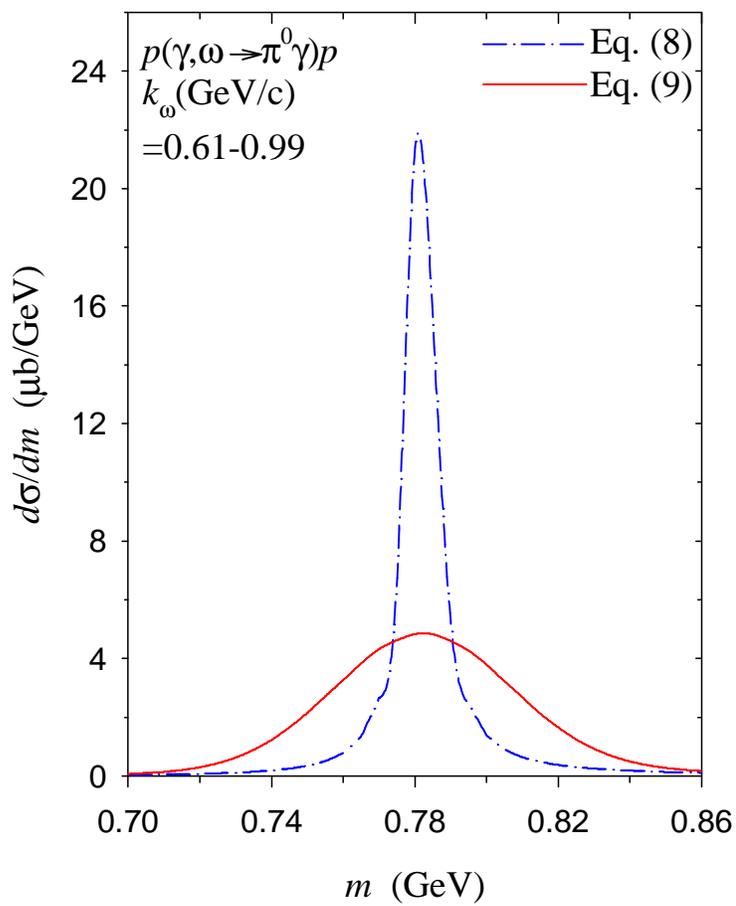
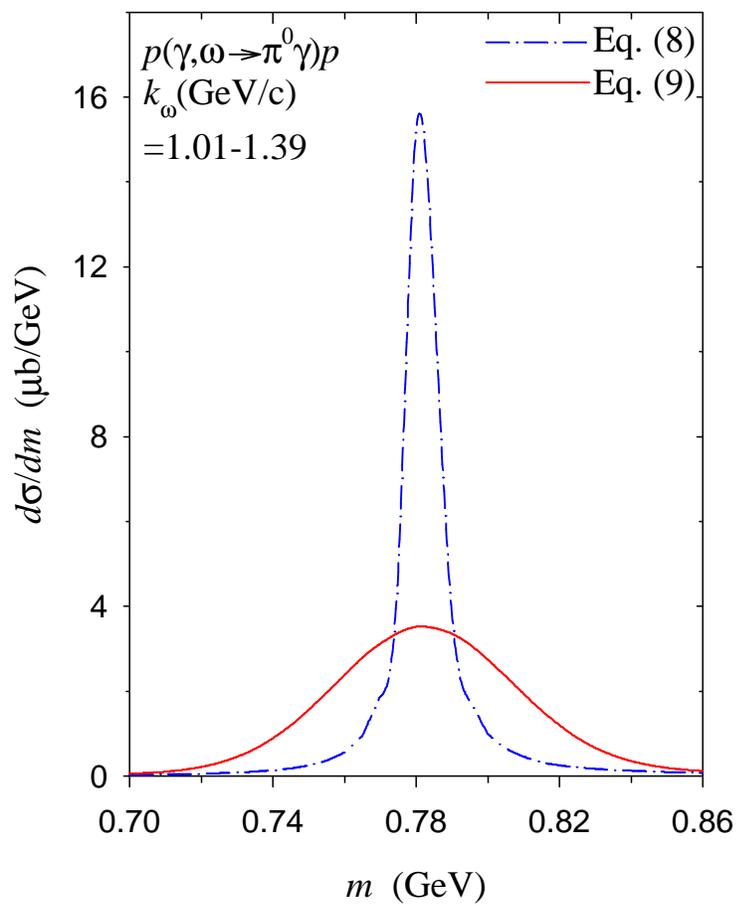
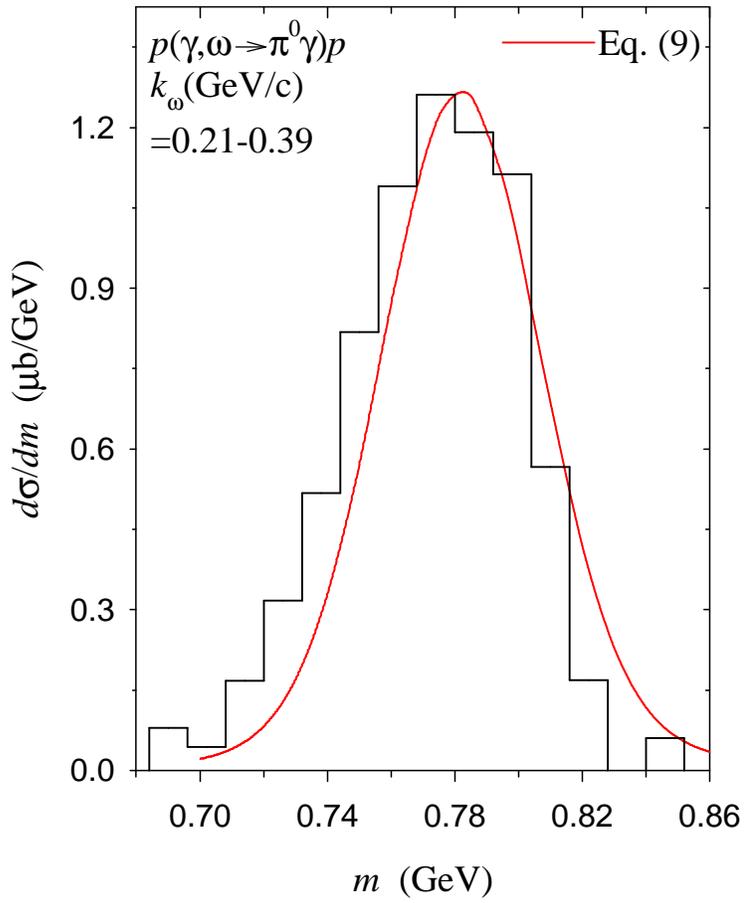
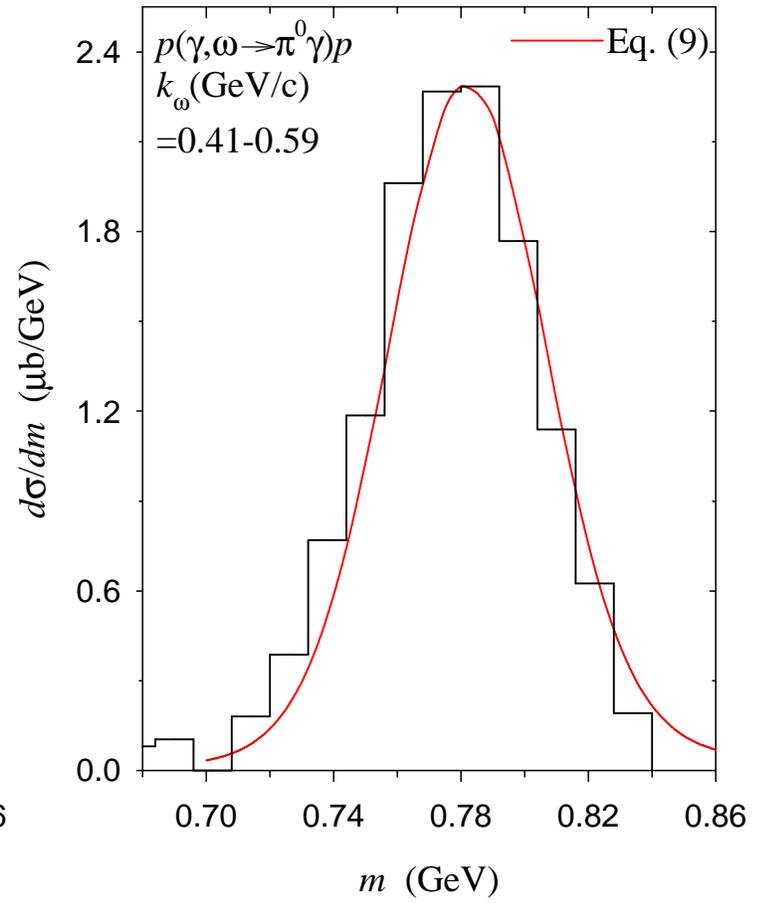
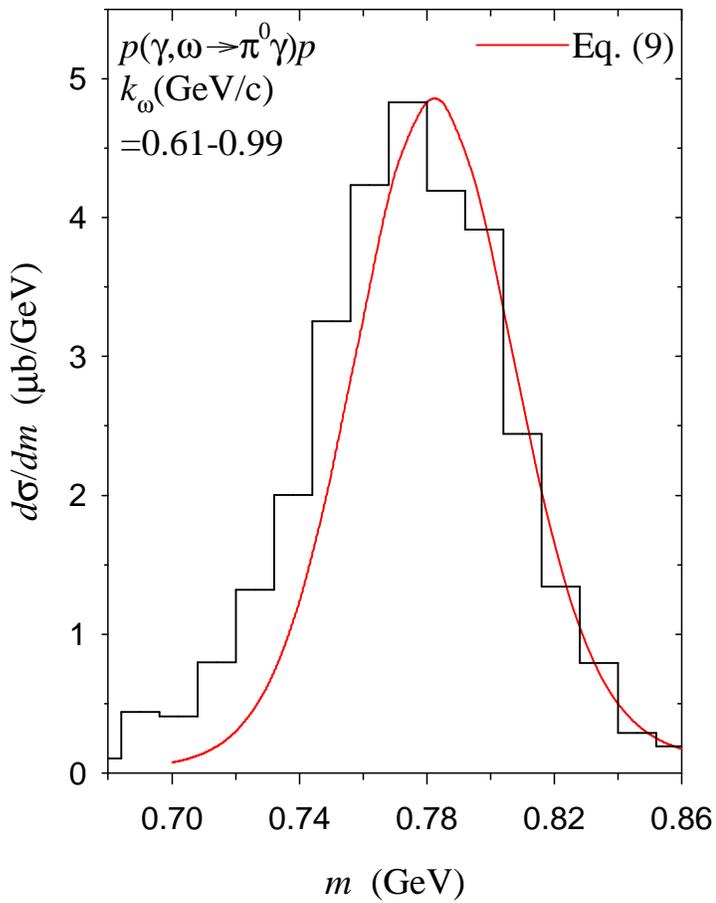
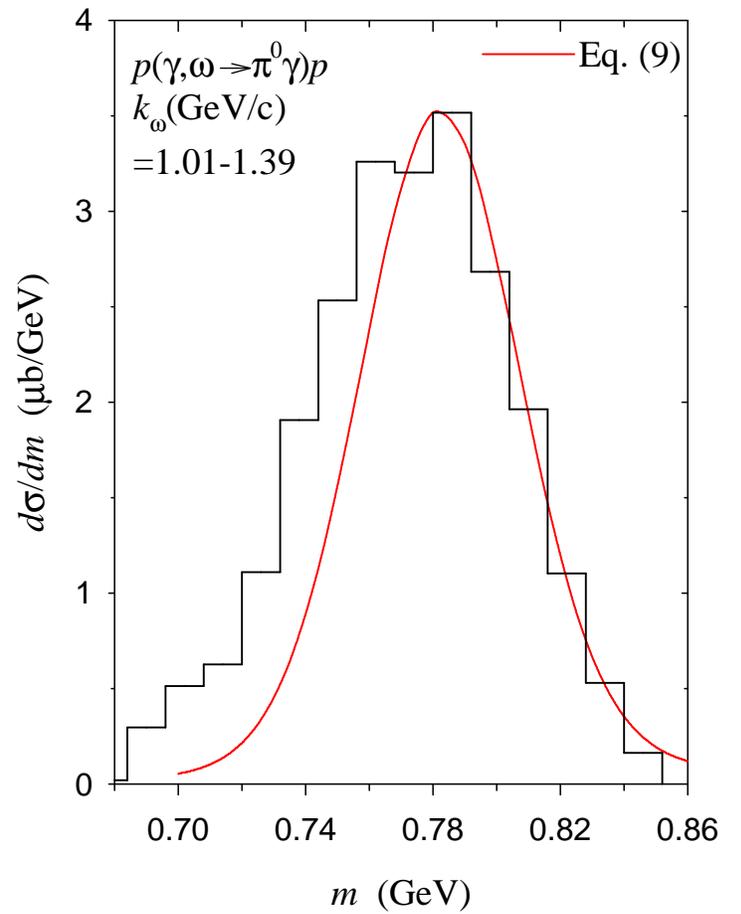
\end{document}